\documentclass{ws-ijmpa}
\usepackage[super,compress]{cite}
\usepackage{graphicx}
\usepackage{xspace}
\usepackage{comment}
\usepackage{url}
\usepackage{hyperref}
\hypersetup{colorlinks,citecolor=black,filecolor=black,linkcolor=black,urlcolor=blue}

\newcommand*{\sqs}{\ensuremath{\sqrt{s}}\xspace}
\newcommand*{\sqsn}{\ensuremath{\sqrt{s_{\rm NN}}}\xspace}

\newcommand*{\pT}{\ensuremath{p_\mathrm{T}}\xspace}
\newcommand*{\mT}{\ensuremath{m_\mathrm{T}}\xspace}

\newcommand*{\MeV}{\ensuremath{\mathrm{MeV}}\xspace}

\newcommand*{\Lz}{\ensuremath{\Lambda^0}\xspace}

\newcommand*{\phiz}{\ensuremath{\phi^0}\xspace}
\newcommand*{\Dz}{\ensuremath{{\rm D}^0}\xspace}
\newcommand*{\Kp}{\ensuremath{{\rm K}^+}\xspace}
\newcommand*{\Kstar}{\ensuremath{{\rm K}^{0\ast}}\xspace}
\newcommand*{\Kzs}{\ensuremath{{\rm K}^0_{\rm s}}\xspace}
\newcommand*{\Dp}{\ensuremath{{\rm D}^+}\xspace}
\newcommand*{\Dstar}{\ensuremath{{\rm D}^{\ast+}}\xspace}
\newcommand*{\pip}{\ensuremath{\pi^+}\xspace}
\newcommand*{\pr}{\ensuremath{\rm p}\xspace}

\newcommand{\dd}{\ensuremath{\mathrm d}\xspace}

\newcommand*{\Teq}{\ensuremath{T_{\rm eq}}\xspace}
\newcommand*{\qeq}{\ensuremath{q_{\rm eq}}\xspace}

\usepackage{soul}

\usepackage{lineno}
\usepackage{orcidlink}
\newcommand{\orcidA}{\orcidlink{0000-0002-2420-7650}} 
\newcommand{\orcidB}{\orcidlink{0000-0003-2849-0120}} 
\newcommand{\orcidC}{\orcidlink{0000-0001-9223-6480}} 
\newcommand{\orcidD}{\orcidlink{0000-0003-3706-5265}} 

\begin{document}
\markboth{L. Gyulai, G. Bíró, R. Vértesi, G. G. Barnaföldi}{Thermodynamical properties of charmed and light hadron systems}

\catchline{}{}{}{}{}
%
%%%%%%%%%%%%%%%%%%%%%%%%%%%%%%%%%%%%%%%%%%%%%%%%%%%%%%%%%%%%%%%%%%%%

\title{Evolution of the hot dense matter at LHC energies through light and heavy-flavor hadrons using non-extensive thermodynamics\\
\vspace{0.2cm}
\textnormal{\it{Dedicated to the memory of J. D. Bjorken}}}

\author{László Gyulai\orcidA$^{\ast,\dagger}$}
\address{gyulai.laszlo@wigner.hun-ren.hu}

\author{Gábor Bíró\orcidB$^{\ast,\ddagger}$}
\address{biro.gabor@wigner.hun-ren.hu}

\author{Róbert Vértesi\orcidD$^{\ast}$}
\address{vertesi.robert@wigner.hun-ren.hu}

\author{Gergely Gábor Barnaföldi\orcidC$^{\ast}$}
\address{barnafoldi.gergely@wigner.hun-ren.hu}

\address{$^{\ast}$HUN-REN Wigner Research Centre for Physics, 29--33 Konkoly--Thege Mikl\'os \'ut,\\ H-1121 Budapest, Hungary \\ 
$^{\dagger}$Budapest University of Technology and Economics, 3 Műegyetem rkp.,\\ H-1111 Budapest, Hungary \\
$^{\ddagger}$E\"otv\"os Lor\'and University, Institute of Physics and Astronomy, 1/A P\'azm\'any P\'eter s\'et\'any,\\ H-1117  Budapest, Hungary\\}

\maketitle

\begin{abstract}
We investigate the formation and evolution of hot systems comprising charmed and light hadrons using non-extensive thermodynamics. We analyze data from pp, p–-Pb, and Pb–-Pb collisions at center-of-mass energies ranging from \sqsn = 2.76~TeV to 13~TeV, measured by the CERN LHC ALICE experiment. The hadron species examined include charged pions and kaons, \Kzs, (anti)protons, $\phi$ mesons, \Lz hyperons, and D mesons. Employing our previously established methods, we determine the common Tsallis parameters \Teq and \qeq for each hadron type. While charm comes from earlier than light hadrons, we see that \Teq is ordered by mass, reflecting a similar ordering in the time-scale relevant for the spectrum. Our results also allow for constraining the heat capacity of the system. The current analysis thus enhances our understanding of hadron production dynamics and thermal properties in high-energy collisions.

\keywords{ultrarelativistic hadron-hadron collisions; non-extensive thermodynamics; charm; hadronization}
\end{abstract}

\ccode{PACS numbers: 25.75.Ag, 13.85.-t, 12.38.Mh, 12.38.-t, 64.70.qd, 14.40.Lb}

\section{Introduction}	

In the theory of quantum chromodynamics (QCD) partons (quarks and gluons) at low energies are expected to exist in color singlet states, confined into hadrons. In the extreme conditions of high-energy heavy-ion collisions, hadronic matter deconfines and forms a new phase, called the quark--gluon plasma (QGP). Quarks and gluons in the QGP acquire color degrees of freedom by not being required to be in the color singlet state anymore. Such a state of matter is believed to have existed in the Universe microseconds after the Big Bang. Studying its properties thus provides a glimpse into the primordial Universe. Over the last decades, experiments at modern collider facilities, such as the CERN's Large Hadron Collider (LHC) and the BNL's Relativistic Heavy Ion Collider (RHIC), played a crucial role in expanding our knowledge of the QGP by measuring many of its fundamental properties~\cite{Ruan:2010vc,Bala:2016hlf}. These experiments observed collective phenomena in high-energy heavy-ion collisions, attributed to the presence of a strongly coupled QGP~\cite{Heinz:2013th}. Recently, collectivity of the hadronic final state has also been observed in smaller collision systems, such as proton--nucleus (p--A)~\cite{CMS:2012qk,ALICE:2012eyl} or even proton--proton (pp)~\cite{CMS:2010ifv,ATLAS:2015hzw}, with high final-state multiplicity, and even in low final-state multiplicity events~\cite{ALICE:2023ulm}. It is still an open question, whether collectivity in small systems can be attributed to the presence of small QGP droplets, or is rather a result of complex vacuum-QCD effects~\cite{Schlichting:2016xmj}. Describing the observations within a unified framework poses a challenge for theories.

In high-energy collisions, light-flavor hadrons (including also the strange hadrons) and heavy-flavor hadrons carry different information. The bulk of the light-flavor hadrons cannot provide detailed information about the earlier stages of a collision, as their kinematics are primarily determined by the conditions present at the freeze-out stage~\cite{Braun-Munzinger:2003pwq}. This is, however, not the case with heavy-flavor hadrons, which are formed by the fragmentation of heavy quarks ($c$ and $b$). These are, in turn, produced at the early stage of the collision. As the lifetime of the heavy quarks is rather long, and their annihilation cross section is negligible, they survive throughout the collision and may serve as sensitive probes of the 
formation and the evolution of the system~\cite{Andronic:2015wma}. The detection of heavy-flavor hadrons posed a challenge for a long time, however, with the recent rapid development of detector capabilities, the heavy-flavor mesons are reconstructed with high precision down to very low (for D$^0$, even zero) transverse momentum\cite{ALICE:2017olh}.

The reconstructed spectra of identified light- and heavy-flavor particles produced in high-energy collisions contain soft, thermally produced particles, as well as particles formed in hard QCD processes. The unification of these two contrasting regions became possible within the non-extensive Tsallis\,--\,Pareto statistical framework\cite{Tsallis:1987eu}. The so-called Tsallis-thermometer\cite{Biro:2020kve} has been shown to be a sensitive tool to learn about the system size, as well as the comparative production timelines of light- and heavy-flavor hadrons\cite{Gyulai:2024dkq}. In the current work, we expand the possibilities of the application of non-extensive thermodynamical principles by studying the relations between different hadron species.

\section{Analysis and results}

In our study, we analyzed data of pp, p--Pb, and Pb--Pb collisions with center-of-mass energies between \sqsn=2.76 TeV and 13 TeV from the measurements of the LHC ALICE experiment. The data included production yields of charged pions\cite{ALICE:2016dei, ALICE:2013wgn, ALICE:2018pal, ALICE:2013mez, ALICE:2019hno} (\pip), kaons\cite{ALICE:2016dei, ALICE:2013wgn, ALICE:2018pal, ALICE:2013mez, ALICE:2019hno, ALICE:2013cdo, ALICE:2016sak, ALICE:2017ban, ALICE:2016fzo, ALICE:2019avo} (\Kp, \Kstar, \Kzs), protons\cite{ALICE:2016dei, ALICE:2013wgn, ALICE:2018pal, ALICE:2013mez, ALICE:2019hno} (\pr), \phiz mesons\cite{ALICE:2018pal, ALICE:2016sak,ALICE:2017ban}, \Lz baryons\cite{ALICE:2013wgn, ALICE:2019avo, ALICE:2013cdo, ALICE:2016fzo} and D mesons\cite{ALICE:2017olh,ALICE:2019nxm,ALICE:2019fhe,ALICE:2012ab} (\Dz, \Dstar, \Dp), as well as the corresponding charge conjugates.
We analyzed the above datasets with the non-extensive thermodynamical framework by generally following the methods described in our previous works\cite{Biro:2020kve, Gyulai:2024dkq}. 
The mid-rapidity transverse-momentum (\pT) spectra of identified hadrons measured at each collisional system and energy were fitted with the Tsallis--Pareto distribution\cite{Tsallis:1987eu,Biro:2020kve}
\begin{equation}
      \left.\frac{\dd^2N}{2\pi \pT \dd \pT \dd y}\right|_{y\approx0} = A \mT \left[1+\frac{q-1}{T}(\mT-\mu) \right]^{-\frac{q}{q-1}}.
  \label{eq:TS}
\end{equation}
Here the free parameters are the normalization factor $A= gV/(2\pi)^3$ containing the volume $V$ and degeneracy factor $g$, the Tsallis temperature $T$, and the non-extensivity parameter $q$. The transverse mass is $\mT=\sqrt{\pT^2+m^2}$ and the chemical potential $\mu$ is approximated with the rest mass of the given hadron, $m$.
As it has been shown earlier\cite{Biro:2020kve}, $T$ and $q$ from light-flavor hadron spectra exhibit a scaling behavior in charged-hadron multiplicity, as well as in collision energy. This scaling results in a grouping of the Tsallis parameters on the $T$---$q$ diagram towards low event multiplicities, around a point denoted by $\Teq$---$\qeq$, which corresponds to the edge case of a very-low-multiplicity system.
Previously, the \Teq and \qeq values had been treated as common for all light-flavor hadrons\cite{Biro:2020kve}, and it was later found that the heavy-flavor D mesons converge to different values\cite{Gyulai:2024dkq}. In the current study, we do not assume such a commonality and determine the \Teq and \qeq values separately for various hadron species. This was made possible by recent high-precision data, and is achieved first by fitting the $T$---$q$ points of each dataset for the given particle species with a line, consequently filling the slope and displacement values into an $E$---$E\delta^2$ diagram\cite{Biro:2020kve, Gyulai:2024dkq}. By fitting the points in the $E$---$E\delta^2$ diagram with a line we obtain the \qeq value (slope) and \Teq value (displacement) for each of the hadron species. The statistical uncertainties were propagated during these fits. 
Note that while the former studies used data both from RHIC and the LHC, here we focus exclusively on the results from the LHC to ensure that the fundamental thermodynamical properties such as heat capacity are stable in the energy regime under investigation\cite{Basu:2016ibk}.

\subsection{Mass hierarchy}

\begin{figure}[htb]
\centerline{\includegraphics[width=0.75\textwidth]{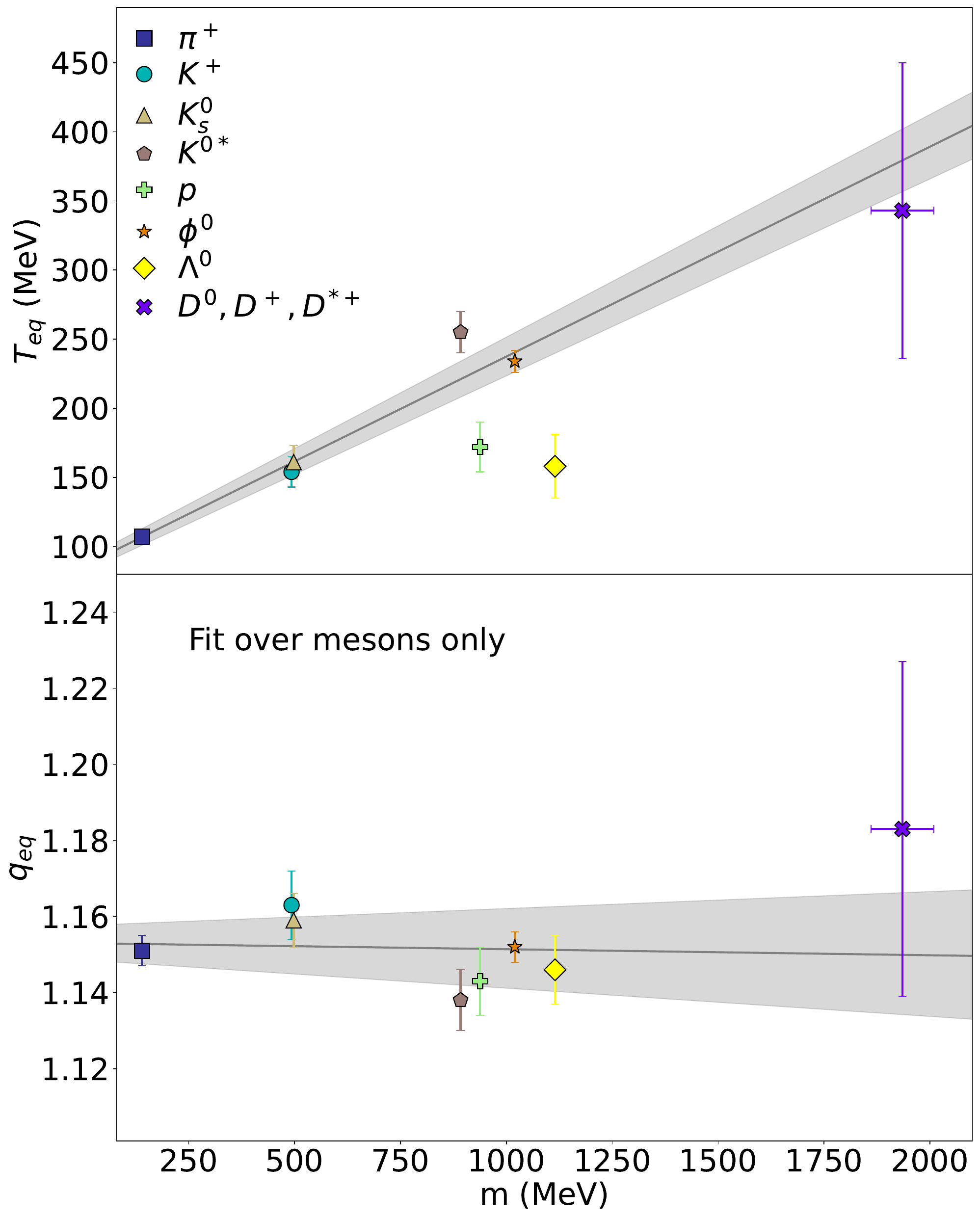}}
\caption{\label{fig:TQvsM}\Teq and \qeq parameters of different hadron species as a function of the invariant mass of the hadrons. The solid lines represent fits over the points corresponding to the mesons, with the gray bands representing the uncertainty range corresponding to one standard deviation.}
\end{figure}

The obtained \Teq and \qeq values are summarized in Table~\ref{tab:qeqTeq}. 
The values are also visualized in Fig.~\ref{fig:TQvsM}, which shows the \Teq and \qeq parameters as a function of the hadron masses.
\begin{table}[htb]
\tbl{Summary of \Teq and \qeq values for different particle species. (Charge conjugates are not marked explicitly.)}{
\begin{tabular}{@{}ccccc@{}} \toprule
Hadron & Quark content & Mass, $m$ & \Teq & \qeq \\
\ & \ & (MeV)&(MeV) &  \\ \colrule
\pip & $u\overline{d}$ & 140 & $107\pm4$ & $1.151\pm0.004$\\
\Kp & $u\overline{s}$ & 493 & $154\pm11$ & $1.163\pm0.009$\\
\Kzs & $\frac{d\overline{s}+\overline{d}s}{\sqrt{2}}$ & 498 & $161\pm12$ & $1.159\pm0.007$\\
\Kstar & $d\overline{s}$ & 892 & $255\pm15$ & $1.138\pm0.008$\\
\pr & $uud$ & 938 & $172\pm18$ & $1.143\pm0.009$\\
\phiz & $s\overline{s}$ & 1020 & $234\pm8$ & $1.152\pm0.004$\\
\Lz & $uds$ & 1115 & $158\pm23$ & $1.146\pm0.009$\\
\Dz, \Dp, \Dstar & $c\overline{u}$, $c\overline{d}$ & 1861--2010 & $343\pm107$ & $1.183\pm0.044$\\
\botrule
\end{tabular} \label{tab:qeqTeq}
}
\end{table}
The common Tsallis temperature \Teq increases with the mass of the mesons, and the trend determined by mesons is well described by a linear fit 
\begin{equation}
\Teq=(0.15\pm0.01) \times m + (85\pm 5)\ \MeV
\end{equation}
in $c\equiv 1$ units. Note that the baryons do not participate in the trend of mass-ordered formation times for mesons, as their \Teq parameter is significantly smaller compared to the mesons with similar masses, hinting at later formation times for baryons compared to mesons. 
The mass dependence is well in line with earlier observations of mass hierarchy from the Tsallis temperatures obtained directly from spectrum fits\cite{Biro:2017arf,Biro:2017eip}. While the emerging picture does not support the common logarithmic mass dependence of mesonic and baryonic Tsallis temperatures\cite{Shen:2019zgi}, more detailed baryonic measurements are needed to draw firm conclusions.

All the light-flavor mesons have similar non-extensivity parameters \qeq, varying between 1.15 and 1.16, while the baryons have a slightly lower value of around 1.14. Although a significant difference can be observed between the \qeq values of unified RHIC and LHC light-flavor and D meson data\cite{Gyulai:2024dkq}, the current precision of LHC-only D-meson data does not support any rising trend. A linear fit to mesons yields

\begin{equation}
\qeq =(-1.6 \pm 6.1 ) \cdot 10^{-6} \, \MeV^{-1} \times m + 1.153 \pm 0.004,
\end{equation}
 which is consistent with a constant.

\subsection{Spectrum formation times}

\begin{figure}[htb]
\centerline{\includegraphics[width=0.8\textwidth]{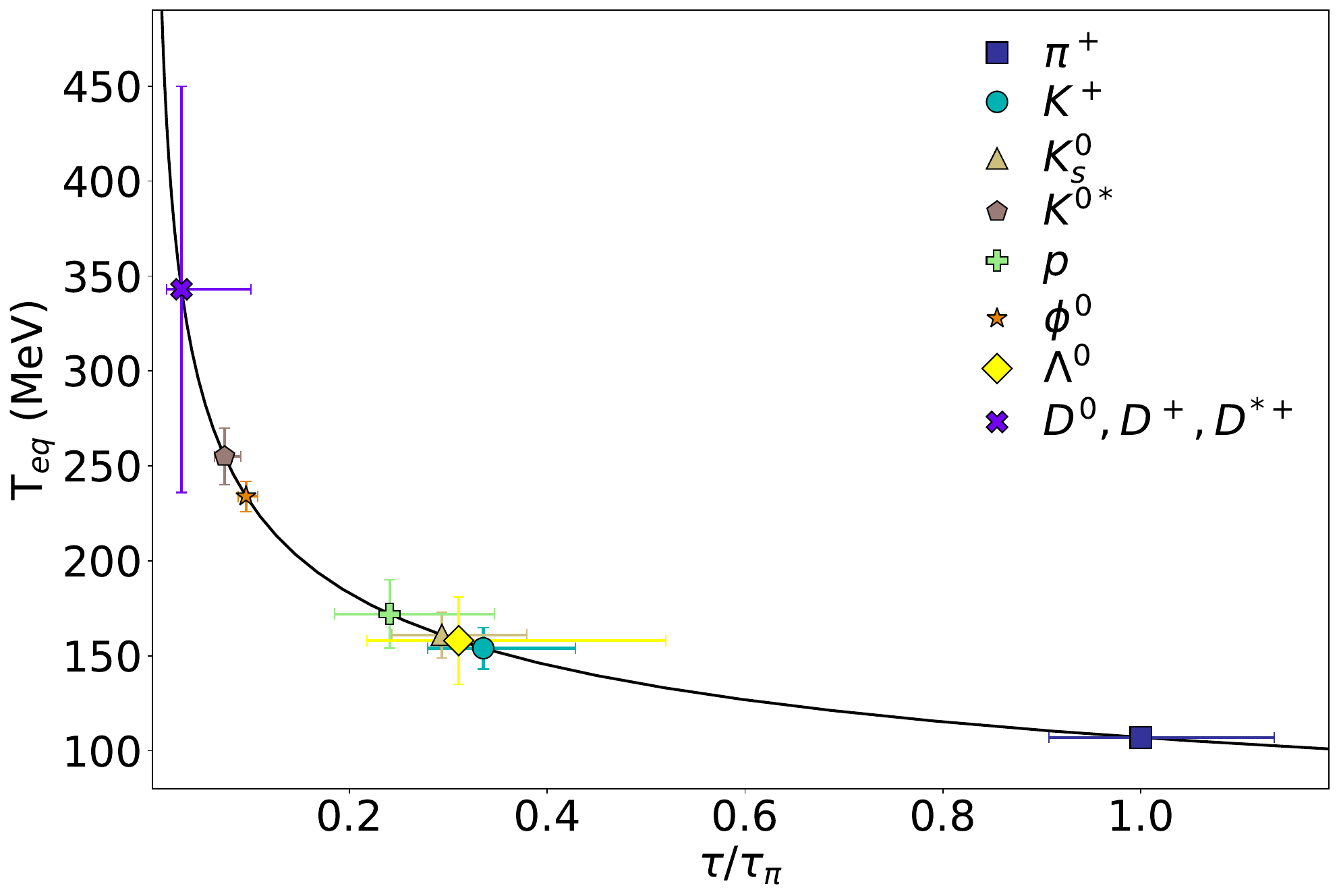}}
\caption{\label{fig:BjorkenTime}Relative formation proper times for each hadron species within the Bjorken model. The proper time is expressed in units relative to that of the pion, $\tau_\pi$.}
\end{figure}

To estimate the formation times of hadron spectra, we postulated a Bjorken-type expansion\cite{Bjorken:1982qr}. This scenario provides a simplified framework for understanding the dynamics of matter created in ultra-relativistic hadron collisions\cite{Heinz:2004qz}. In this model, the system is assumed to undergo a boost-invariant expansion along the beam axis, where the proper time evolution of energy density and temperature follows a scaling. The relation between temperature and proper time can be expressed as\cite{Magas:2017nuw}
\begin{equation}\label{eq:tauT}
    \tau = \tau_0 \left( \frac{T_0}{T} \right)^3 \ ,
\end{equation}
where $\tau_0$ and $T_0$ correspond to a given initial proper time and temperature, respectively. Since the Bjorken expansion is independent of the thermodynamical picture, the non-extensive framework can be utilized\cite{Gyulai:2024dkq}. The formation proper times of each hadron species were estimated with respect to the proper time of pions in order to avoid defining the reference points $T_0$ and $\tau_0$. The results are presented in Fig.~\ref{fig:BjorkenTime}.
It can be observed that the meson formation times are ordered by mass, as the spectra of heavier mesons correspond to an earlier formation time, while spectra of lighter mesons are formed later. Interestingly, heavy-flavor D mesons are no exception from this trend. Baryons are formed later than mesons with similar masses. According to our results, pion spectra are formed at a substantially later time compared to all the other hadrons. The proper time associated with the formation of pions is approximately 3 times larger than charged kaons, while D meson spectra are formed a further order of magnitude earlier than those of kaons ($\tau_{\rm K} / \tau_\pi \approx 0.34$, $\tau_{\rm D} / \tau_\pi \approx 0.03$). These results also agree with our earlier estimation between the formation proper times of heavy-flavor and light-flavor hadron spectra (not differentiated by species), $\tau_{\rm HF} / \tau_{\rm LF} \approx 0.18$\cite{Gyulai:2024dkq}.

\subsection{Non-extensivity parameter and heat capacity}

For systems with a fluctuating number of particles, the non-extensivity parameter $q$ can be generally written as\cite{Biro:2014cka}

\begin{equation}\label{eq:qdef}
    q=1-\frac{1}{C}+\delta^2 \ ,
\end{equation}

\noindent where $C$ is the heat capacity of the system, while $\delta^2$ expresses the temperature or multiplicity fluctuations in the events ($\Delta T^2 / \langle T \rangle^2$ and $\Delta n^2 / \langle n \rangle^2$ respectively). The values of $\delta^2$ are obtained from fits on the $q$---$T$ points of each hadron species during the evaluation of the \qeq and \Teq parameters\cite{Gyulai:2024dkq}. The $\delta^2$ values increase with the size of the collision system, therefore we determined a range for each hadron species, where the lower boundary corresponds to pp, and the higher boundary to Pb--Pb collisions. 

Several attempts have been made to estimate the heat capacity of the hot hadronic system. 
Based on temperature fluctuations, the specific heat capacity $c_V = C/n$ corresponding to a constant volume was estimated at SPS energies to be $c_V=60 \pm 100$, assuming $\langle T \rangle=180$ MeV\cite{Korus:2001au}. In this study, the Tsallis nature of the \pT distribution was considered in the temperature fluctuations, but not in the derivation of the heat capacity. Another work based on the lattice QCD calculations estimates $c_V/T^3\approx15$ in case of QGP and $c_V/T^3\approx21$ for the ideal gas limit\cite{Gavai:2004se}.
More recent estimates on the specific heat in Au--Au and Cu--Cu collisions at RHIC, based on the Boltzmann-Gibbs statistics, yielded $c_V\approx1-2$\cite{Basu:2016ibk}. 
Note that the \sqs dependence of the heat capacity is saturated in the ultra-relativistic limit\cite{Basu:2016ibk}, therefore this estimate is also valid at LHC energies.
However, this approach relies on extensive thermodynamics where $q = 1$, therefore Eq.~\eqref{eq:qdef} reduces to $C = 1/\delta^2$. Non-extensive thermodynamics can provide corrections to this estimate. The Tsallis--Pareto fits to the spectra yield $q$ values significantly above unity, meaning that the specific heat of the system is also expected to be higher than in the case of Boltzmann-Gibbs distribution.
Another study based on the non-extensive framework obtained the value of specific heat to be around $c_V\approx 1-4$ depending on the value of $q$ and $T$\cite{Sahu:2021cdl}.

Here we use a different approach. We estimate the specific heat based on the common non-extensivity parameter \qeq, corresponding to the low-multiplicity limit. By substituting the \qeq and $\delta^2$ values for each particle species in Eq.~\eqref{eq:qdef}, we can obtain the specific heat of the very-low-multiplicity systems corresponding to the \Teq and \qeq values. The results across the different particle species are consistent with each other, yielding a lower boundary for the specific heat, $C\gtrapprox5$. 
Considering the system sizes this is consistent with the values of the previous non-extensive study\cite{Sahu:2021cdl}.
Due to the hyperbolic dependence between \qeq and $C$, the upper limit of the specific heat cannot be well estimated by the current precision of the available data. 
This can be interpreted as follows. For light mesons (specifically pions and kaons) produced in larger collision systems, $\delta^2 \approx q-1$, leading to large $C$. This implies a largely thermalized system. For heavier mesons including the heavy-flavor D mesons, which correspond to earlier spectrum formation times, the relative fluctuations are larger, which leads to smaller heat capacity and imply a strongly non-extensive system.
Using the non-extensive thermodynamical approach with future more precise data, especially on rarer hadron species, will help addressing time development of the heat capacity $C$ by exploiting its connection to the time dependence of the non-extensivity parameter $q$.

\section{Conclusions}

We analyzed data from pp, p–-Pb, and Pb–-Pb collisions at center-of-mass energies ranging from \sqsn = 2.76~TeV to 13~TeV, measured by the LHC ALICE experiment, to investigate the formation and evolution of hot systems comprising charmed and light hadrons using non-extensive thermodynamics. This work builds on our previous paper\cite{Gyulai:2024dkq}, which focused on differences in thermodynamical properties between heavy- and light-flavor hadron data at RHIC and LHC energies. In the present study, we extend our analysis to the individual evaluation of various hadron species measured at a single facility.

\begin{itemize}

    \item We determined the common Tsallis parameters \Teq and \qeq for  charged pions and kaons, \Kzs, (anti)protons, $\phi$ mesons, \Lz hyperons, and D mesons separately. We found that \Teq for all the mesons, both light and charm, follow a mass ordering, where \Teq linearly increases with mass. Baryons have a slightly lower \Teq than mesons with the same mass. On the other hand, LHC data does not show a significant ordering of \qeq with the mass of different mesons, as a linear fit yields, consistent with no slope.
    
    \item Using Bjorken expansion as an assumption, we determined the spectrum formation proper times for each hadron species relative to each other. As expected, the proper times are mass ordered and pion spectra are formed at a substantially later time compared to all the other hadrons ($\tau_{\rm K} / \tau_\pi \approx 0.34$ for charged kaons, $\tau_{\rm D} / \tau_\pi \approx 0.03$ for D mesons).

    \item We estimated the heat capacity of the system based on the common non-extensivity parameters \qeq and relative fluctuations $\delta^2$ corresponding to each hadron species. The results yield a system-wide lower boundary, $C\gtrapprox5$. The upper limit corresponding to light mesons implies a largely thermalized system, while heavier hadrons are strongly non-extensive regardless of the system size. This is consistent with the earlier formation of heavier hadron spectra.
   
\end{itemize}

With the methods outlined above, future precise event-multiplicity-dependent identified-hadron measurements at the LHC, especially in the heavy-flavor sector will allow for the exploration of thermodynamical properties of the hot, strongly interacting matter in even more detail. Similar measurements at RHIC, on the other hand, can pin down the extent of collision-energy-dependence of formation times and heat capacity.
    
\section*{Acknowledgments}

This work has been supported by the NKFIH grants OTKA FK131979 and K135515, as well as by the 2021-4.1.2-NEMZ\_KI-2024-00031, 2021-4.1.2-NEMZ\_KI-2024-00033 and 2021-4.1.2-NEMZ\_KI-2024-00034 projects. The authors acknowledge the research infrastructure provided by the Hungarian Research Network (HUN-REN) and the Wigner Scientific Computing Laboratory.

\bibliographystyle{ws-ijmpa}
\bibliography{references}

\end{document}